%% file: Multi_axion_EDE_letter.tex
\documentclass[twocolumn,showkeys,showpacs,preprintnumbers,prd,superscriptaddress,nofootinbib]{revtex4-1}
\usepackage{graphicx,epsf,bm,amsmath,amsfonts,amssymb,epstopdf,natbib,hyperref,color,verbatim,multirow,bm,siunitx}
\hypersetup{colorlinks=true,urlcolor=blue,citecolor=blue,linkcolor=blue,menucolor=blue,anchorcolor=blue,filecolor=blue}
\usepackage{appendix}

\newcolumntype{?}{!{\vrule width 3pt}}
\date{\today}

\begin{document}

\title{Double the axions, half the tension: \\
multi-field early dark energy eases the Hubble tension}

\author{Marco Bella}
\email{marco.bella@studenti.unitn.it}
\affiliation{Department of Physics, University of Trento, Via Sommarive 14, 38123 Povo (TN), Italy}

\author{Vivian Poulin}
\email{vivian.poulin@umontpellier.fr}
\affiliation{Laboratoire Univers \& Particules de Montpellier, CNRS \& Universit\'{e} de Montpellier, 34095 Montpellier, France \looseness=-1}

\author{Sunny Vagnozzi}
\email{sunny.vagnozzi@unitn.it}
\affiliation{Department of Physics, University of Trento, Via Sommarive 14, 38123 Povo (TN), Italy}
\affiliation{Trento Institute for Fundamental Physics and Applications (TIFPA)-INFN, Via Sommarive 14, 38123 Povo (TN), Italy}

\author{Lloyd Knox}
\email{lknox@ucdavis.edu}
\affiliation{Department of Physics and Astronomy, University of California, One Shields Avenue, Davis, CA 95616, USA}

\begin{abstract}
\noindent We show that the strong constraints placed by Planck NPIPE Cosmic Microwave Background (CMB) data on axion-like early dark energy (EDE) are significantly alleviated in models with multiple fields. We find a $1.5\sigma$ residual tension with the Local Distance Network value of $H_0$ in a 2-field model, with no improvement beyond two fields, and a best-fit value of $H_0$ $\sim 1.4\sigma$ larger than in the 1-field case. The second field improves the fit to high-$\ell$ CMB data, where 1-field EDE is most strongly disfavored, and suggests modifications to the pre-recombination history over a wider redshift range.
\end{abstract}

\maketitle
\textbf{\textit{Introduction.}} --- Over the past decade, a growing disagreement between low- and high-redshift determinations of the Hubble constant has emerged~\cite{Verde:2019ivm,DiValentino:2020zio,DiValentino:2021izs,Perivolaropoulos:2021jda,Shah:2021onj,Abdalla:2022yfr,DiValentino:2022fjm,Hu:2023jqc,Verde:2023lmm,CosmoVerseNetwork:2025alb}. The value recently reported by the Local Distance Network (H0DN), $H_0=(73.50 \pm 0.81)\,\text{km/s/Mpc}$~\cite{H0DN:2025lyy}, disagrees at more than $7\sigma$ with the value $H_0=(67.24 \pm 0.35)\,\text{km/s/Mpc}$ inferred, assuming the $\Lambda$CDM model, from a combination of the latest Planck, ACT, and SPT Cosmic Microwave Background (CMB) data~\cite{SPT-3G:2025bzu}. Consistently low values of $H_0$ are also inferred from CMB-free dataset combinations~\cite{Cuceu:2019for,Schoneberg:2019wmt}, e.g.\ combining Big Bang Nucleosynthesis information with the latest Baryon Acoustic Oscillation (BAO) data from DESI yields $H_0=(68.50 \pm 0.58)\,\text{km/s/Mpc}$~\cite{DESI:2025zgx}. Attempts to explain the tension via (multiple, independent) systematics~\cite{Efstathiou:2020wxn,Mortsell:2021tcx,Kenworthy:2022jdh,Camarena:2022iae,Wojtak:2022bct,Riess:2023bfx,Giani:2023aor,Wojtak:2024mgg,Freedman:2024eph,Perivolaropoulos:2024yxv,Riess:2025chq,Hogas:2026urs} are facing difficulties, and serious consideration is therefore given to the possibility that the Hubble tension calls for new physics beyond $\Lambda$CDM~\cite{Mortsell:2018mfj,Vagnozzi:2018jhn,Yang:2018euj,Guo:2018ans,Kreisch:2019yzn,Vagnozzi:2019ezj,Visinelli:2019qqu,DiValentino:2019ffd,DiValentino:2019jae,Krishnan:2020obg,Alestas:2020mvb,Jedamzik:2020krr,Sekiguchi:2020teg,RoyChoudhury:2020dmd,Brinckmann:2020bcn,Gao:2021xnk,Marra:2021fvf,Dainotti:2021pqg,Krishnan:2021dyb,Cyr-Racine:2021oal,Anchordoqui:2021gji,Akarsu:2021fol,Ren:2022aeo,Nojiri:2022ski,Schoneberg:2022grr,Banerjee:2022ynv,deSa:2022hsh,Akarsu:2022typ,Lee:2022gzh,Khodadi:2023ezj,Bernui:2023byc,Ben-Dayan:2023rgt,Gomez-Valent:2023hov,Ruchika:2023ugh,Adil:2023exv,Frion:2023xwq,Gomez-Valent:2023uof,Akarsu:2024qiq,Giare:2024ytc,Lynch:2024gmp,Giare:2024smz,Lynch:2024hzh,Toda:2024ncp,Nozari:2024wir,Escamilla:2024xmz,RoyChoudhury:2024wri,Mirpoorian:2024fka,Li:2025owk,Lee:2025yah,Teixeira:2025czm,Wang:2025dzn,Zhang:2025dwu,Kumar:2025obb,Hogas:2025mii,Pantos:2026rpe}.

Because of stringent constraints imposed by BAO and uncalibrated Type Ia Supernovae (SNeIa) measurements, a successful solution to the Hubble tension is required to operate primarily prior to recombination, and reduce the sound horizon~\cite{Bernal:2016gxb,Lemos:2018smw,Aylor:2018drw,Knox:2019rjx,Arendse:2019hev,Cai:2021weh,Keeley:2022ojz,Jiang:2024xnu,Zhou:2025kws,Pedrotti:2025ccw}. The Early Dark Energy (EDE) class of models~\cite{Karwal:2016vyq,Poulin:2018cxd,Smith:2019ihp,Kamionkowski:2022pkx,Poulin:2023lkg,McDonough:2023qcu} has emerged as particularly promising in this context~\cite{Agrawal:2019lmo,Lin:2019qug,Niedermann:2019olb,Berghaus:2019cls,Sakstein:2019fmf,Ye:2020btb,Zumalacarregui:2020cjh,Ballesteros:2020sik,Braglia:2020iik,Ballardini:2020iws,Gogoi:2020qif,Braglia:2020bym,Gonzalez:2020fdy,CarrilloGonzalez:2020oac,Braglia:2020auw,Adi:2020qqf,Karwal:2021vpk,Jiang:2021bab,Gomez-Valent:2021cbe,Poulin:2021bjr,Niedermann:2021vgd,Wang:2022jpo,Smith:2022hwi,Jiang:2022uyg,Gomez-Valent:2022bku,Escudero:2022rbq,Herold:2022iib,Goldstein:2023gnw,Eskilt:2023nxm,Odintsov:2023cli,Sharma:2023kzr,Fu:2023tfo,Pedreira:2023qqt,Khalife:2023qbu,Forconi:2023hsj,Wang:2024jug,Giare:2024akf,Chatrchyan:2024xjj,Jiang:2024nha,Forconi:2025cwp,Piras:2025eip,Stahl:2025czl,Jiang:2025hco,Forconi:2025zzu,Smith:2025grk,Yashiki:2025loj,Toda:2025kcq,Yin:2026gss,Gonzalez-Fuentes:2026rgu,Jhaveri:2026bla}. For instance, the ``axion-like'' EDE model brings the residual tension below $2\sigma$ when confronted with ACT DR6 CMB and DESI DR2 BAO data~\cite{Poulin:2025nfb}, and to $2.3\sigma$ when confronted against ACT, SPT, Planck PR3, and DESI DR2 data~\cite{SPT-3G:2025vyw}. The most significant challenge to the model's success, however, is CMB data from the latest Planck NPIPE release~\cite{Planck:2020olo}: when combined with pre-DESI BAO data, NPIPE data precludes a full solution to the Hubble tension, leaving a residual $3.7\sigma$ tension~\cite{Efstathiou:2023fbn}.

Our goal in this \textit{Letter} is to show that the tension between EDE and Planck NPIPE data is not a failure of EDE itself, but rather a limitation of its simplest, single-field implementation. We introduce a multi-field EDE sector featuring several pseudo-scalar EDE fields with different masses and decay constants. This ``multi-axion'' setup enjoys strong theoretical motivation within the string axiverse context~\cite{Svrcek:2006yi,Arvanitaki:2009fg,Cicoli:2012sz,Visinelli:2018utg,Cicoli:2023opf}. Our setup and results should be understood as a phenomenological proxy for a richer pre-recombination expansion history, expanding upon the successful 1-field EDE model while being consistent at the perturbation level, and providing a physically motivated starting point for future model-independent reconstructions of the pre-recombination expansion history. Importantly, we show that the very tight Planck NPIPE constraints are significantly alleviated if there was \textit{not one, but two} ultralight EDE fields active before recombination. Further, including more fields does not significantly improve over the 2-field model, indicating that \textit{one can not get an arbitrarily successful model by simply increasing the number of EDE fields}.

\textbf{\textit{Analysis setup.}} --- We consider $N_{\text{ax}}=1,2,3$ axion-like EDE fields $\phi_i$, each characterized by its mass $m_i$, decay constant $f_i$, and the following potential~\cite{Poulin:2018cxd}:
\begin{equation}
V_i(\phi_i) = m_i^2 f_i^2 \left [1 - \cos \left ( \frac{\phi_i}{f_i} \right ) \right ] ^n\,.
\label{eq:potential}
\end{equation}
Via a shooting algorithm, we trade $m_i$ and $f_i$ for the phenomenological parameters $z_{c}^i$ and $f_{\text{ax}}^i(z_{c}^i)$, i.e.\ the critical redshift at which the $i$th field becomes dynamical, and the fractional energy density carried by the field at the critical redshift. The $N_{\text{ax}}$ EDE fields are taken to be non-interacting, and their individual perturbations obey the same equations of the standard, single-field EDE model~\cite{Poulin:2023lkg}. We implement these equations in the publicly available \texttt{mAxiCLASS} code,~\footnote{\url{https://github.com/marcobeIIa/mAxiCLASS}} a modified version of the earlier \texttt{AxiCLASS}~\cite{Poulin:2018dzj,Poulin:2018cxd,Smith:2019ihp,Murgia:2020ryi},~\footnote{\url{https://github.com/PoulinV/AxiCLASS}} itself an extension of the \texttt{CLASS}~\footnote{\url{https://github.com/lesgourg/class_public}} Boltzmann solver~\cite{Lesgourgues:2011re,Blas:2011rf}.

Our analysis compares models with $N_{\text{ax}}=1,2,3$ EDE fields. Each of these is allowed to be dynamical in a distinct redshift bin within the overall range $\log_{10}z_c \in [3,4.5]$. For the $N_{\text{ax}}=2$ case we choose $[3,3.75]$ and $[3.75,4.5]$ as redshift bins for $\log_{10}z_c^i$, whereas in the $N_{\text{ax}}=3$ case we choose the non-uniform bins $[3,3.75]$, $[3.75,4.125]$, and $[4.125,4.5]$, for both physical and statistical reasons. On the statistical side, this ensures that the 3-axion case is nested within the 2-axion one, allowing for meaningful model comparison at a later stage (for a co-prime number of fields, this cannot be achieved with uniform binning). On the physical side, addressing the Hubble tension with EDE appears to consistently require a field which is dynamical within $3 \lesssim \log_{10}z_c \lesssim 3.75$, so we choose not to split the first bin further.~\footnote{We checked that splitting this bin does not improve the fit.} In each bin, we vary both the critical fractions $f_{\text{ax}}^i(z_c^i) \in [0,0.3]$ and the critical redshifts $z_c^i$. We fix the potential index in Eq.~(\ref{eq:potential}) to $n=3$, and the fields' initial conditions to $\theta_i \equiv \phi_i/f_i = 2.8$, $\dot{\theta}_i = 0$. The choice $n=3$ is motivated by the fact that cosmological data are largely insensitive to $n \sim 2-5$~\cite{Simon:2023hlp}, and ensures that each field dilutes faster than radiation after becoming dynamical. For single-field EDE, data are known to favor $\theta \sim \pi$~\cite{Smith:2019ihp}, and to reduce the number of free parameters we opt for fixing $\theta_i$ close to this value for all $N_{\text{ax}}$ EDE fields.

We test our multi-field EDE models against some of the latest available cosmological data. Our CMB data includes the Planck 2018 \texttt{Commander} and \texttt{SimAll} likelihoods for low-$\ell$ TT and EE, and the PR3 lensing likelihood~\cite{Planck:2018lbu,Planck:2019nip}. We include the high-$\ell$ TTTEEE \texttt{CamSpec} PR4 likelihood~\cite{Rosenberg:2022sdy}, as it provides the tightest constraints on 1-field EDE~\cite{Efstathiou:2023fbn}. We also include PantheonPlus uncalibrated SNeIa data~\cite{Brout:2022vxf}, as well as DESI DR2 BAO data~\cite{DESI:2025zgx}. Finally, when assessing the residual Hubble tension, we include a Gaussian prior on $H_0=(73.50 \pm 0.81)\,\text{km/s/Mpc}$, informed by the Local Distance Network~\cite{H0DN:2025lyy}, which we refer to as H0DN. We refer to the CMB dataset as ``P'', the combination with SNeIa and BAO data as ``PD'', and the combination of all data including the H0DN prior as ``PDH0''.

We perform a Bayesian analysis via Markov Chain Monte Carlo (MCMC) runs using the \texttt{MontePython} sampler~\cite{Brinckmann:2018cvx,Audren:2012wb}, and analyzing our MCMC chains via the \texttt{GetDist} package~\cite{Lewis:2019xzd}. Bayesian analyses of EDE are known to be susceptible to prior volume effects~\cite{Smith:2020rxx,Herold:2021ksg,Gomez-Valent:2022hkb,Reeves:2022aoi,Herold:2024enb}, which can shift the inferred $H_0$ (towards either larger or smaller values) independently of any genuine change in quality of fit. We therefore complement our MCMC analysis with a frequentist profile likelihood (PL) approach using the \texttt{Procoli} package~\cite{Karwal:2024qpt} to determine both the best-fit points in parameter space, and PL on $H_0$.

Following standard practice in the literature, we quantify the residual tension with the H0DN determination computing the difference in best-fit $\chi^2$ between the PDH0 and PD datasets~\cite{Schoneberg:2021qvd}:
\begin{equation}
\Delta\chi^2_{H_0}=\chi^2_\text{min}(\text{PDH0})-\chi^2_\text{min}(\text{PD})\,,
\end{equation}
with the residual tension being $Q_{\text{DMAP}}=\sqrt{\Delta\chi^2_{H_0}}$ in units of $\sigma$~\cite{Raveri:2018wln}. We assess model preference for a given model ${\cal M}$ with respect to the 1-axion case using the Akaike information criterion (AIC)~\cite{Akaike:1974vps}:
\begin{equation}
\Delta\text{AIC}=\chi^2_{\text{min}}(\mathcal{M})-\chi^2_\text{min}(N_{\text{ax}=1})+2k\,,
\end{equation}
where $k$ is the number of extra parameters in model ${\cal M}$ compared to the $N_{\text{ax}}=1$ case (when ${\cal M}=\Lambda$CDM, $k=-2$), and $\Delta\text{AIC}<0$ indicates that model ${\cal M}$ is favored.

\begin{figure}
\centering
\includegraphics[width=0.9\linewidth]{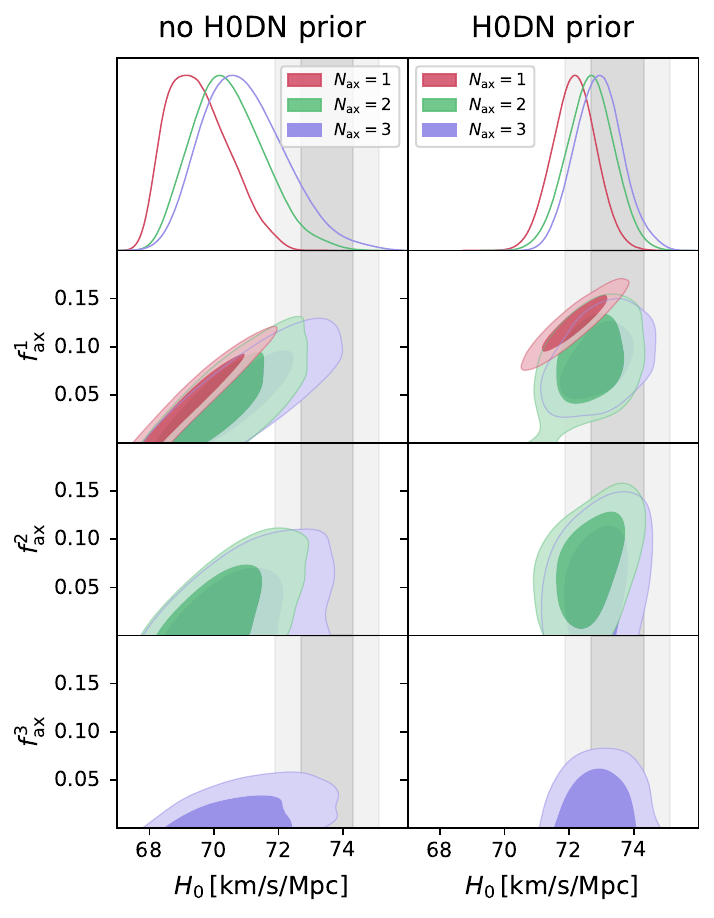}
\caption{Marginalized posteriors for $f_{\text{ax}}^i$ and $H_0$ within the $N_{\text{ax}}=1,2,3$ models, obtained from the PD (left) and PDH0 (right) combinations. The grey bands correspond the the H0DN $68\%$ (dark) and $95\%$ (light) credible intervals.}
\label{fig:123npsd}
\end{figure}

\begin{figure}
\centering
\includegraphics[width=0.9\linewidth]{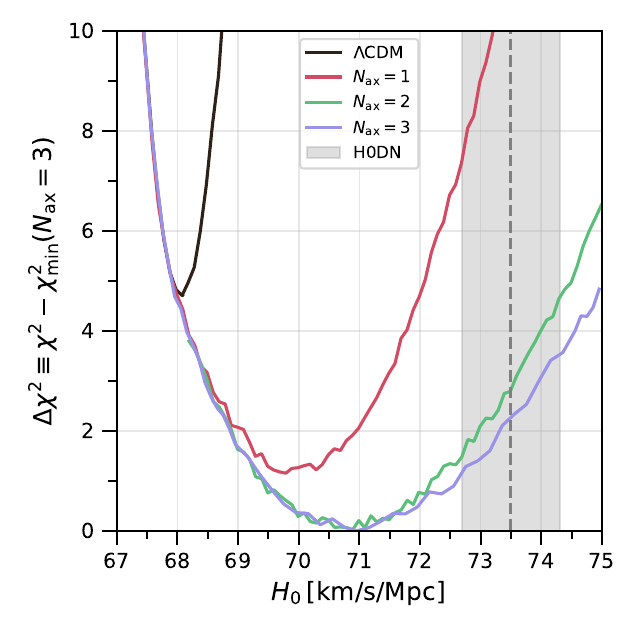}
\caption{$H_0$ profile likelihood in the $\Lambda$CDM and $N_{\text{ax}}=1,2,3$ models, obtained from the H0DN-free PD combination, and normalized to the best-fit $\chi^2$ in the $N_{\text{ax}}=3$ model. The grey band corresponds to the H0DN $68\%$ credible interval.}
\label{fig:profile}
\end{figure}

\begin{figure}
\centering
\includegraphics[width=0.8\linewidth]{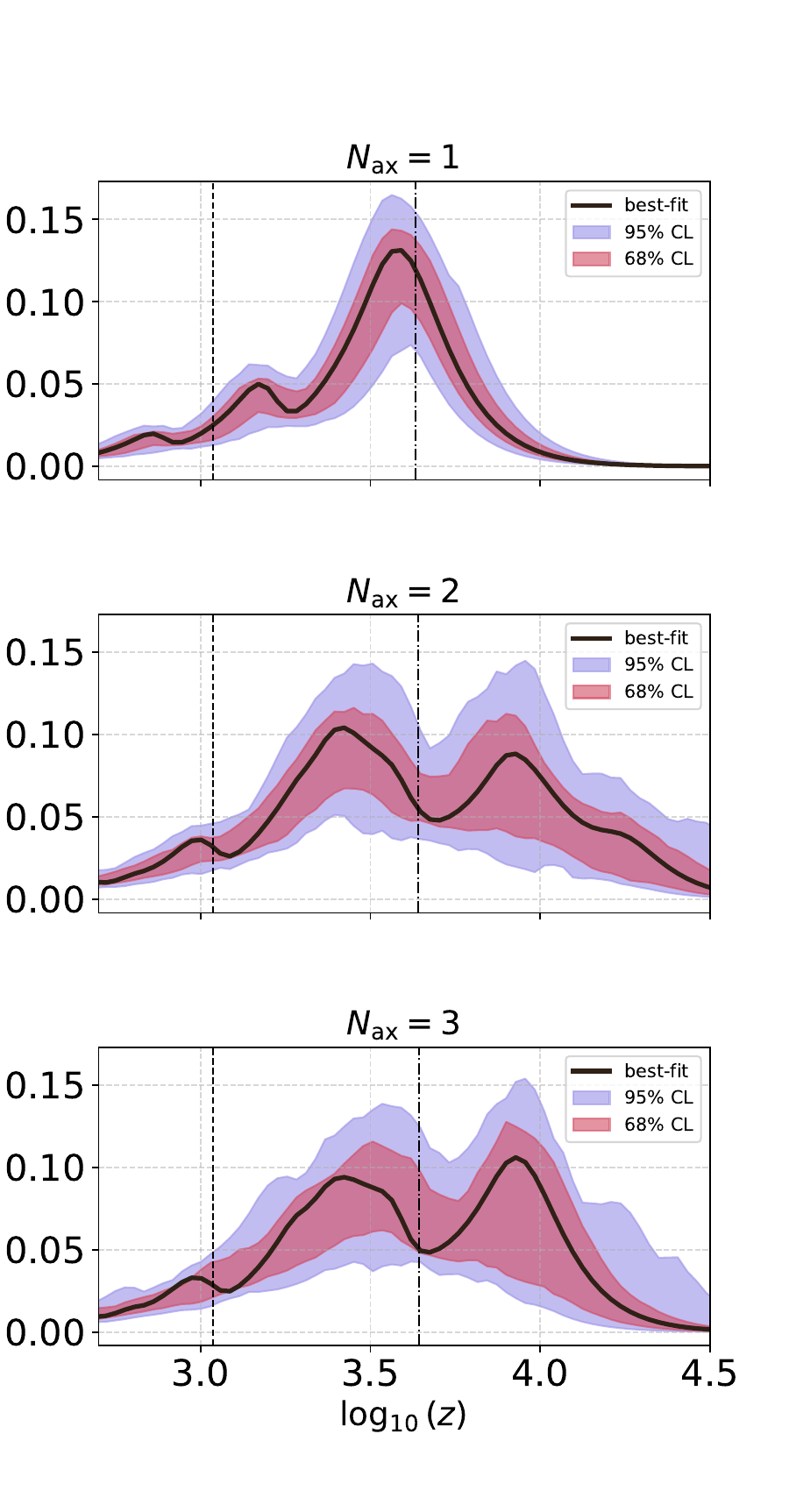}
\caption{Reconstructed posteriors for the total EDE fraction $f_{\text{EDE}}(z)$ for the $N_{\text{ax}}=1,2,3$ models, assuming the PDH0 combination. The dashed (dash-dotted) vertical lines indicate recombination (matter-radiation equality).}
\label{fig:energy_injection}
\end{figure}

\begin{figure}
\centering
\includegraphics[width=0.9\linewidth]{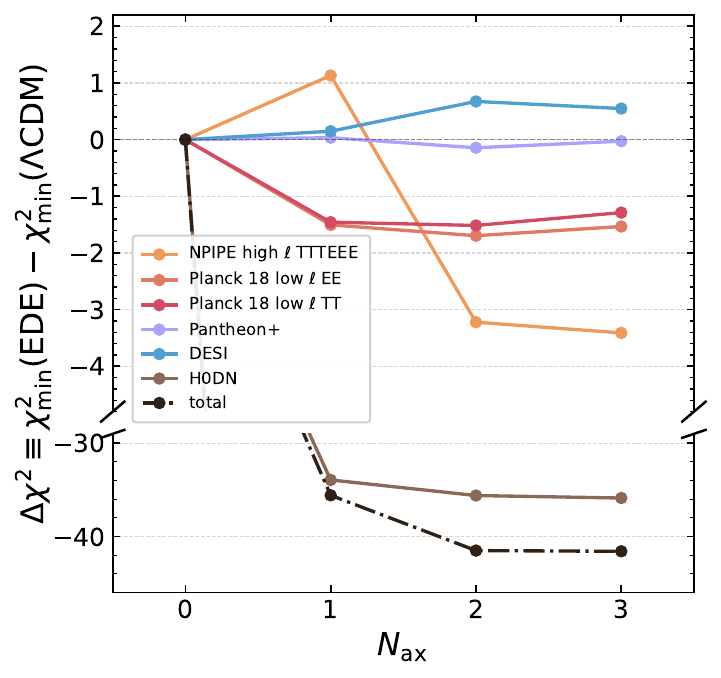}
\caption{Difference in best-fit $\chi^2$ for the $N_{\text{ax}}=1,2,3$ models compared to $\Lambda$CDM, obtained from the PDH0 combination. We show both the total and per-likelihood $\Delta \chi^2$.}
\label{fig:chi2perexp}
\end{figure}

\textbf{\textit{Results.}} --- We report (Bayesian and frequentist) $H_0$ constraints in Tab.~\ref{tab:h0_summary}, and our tension and model comparison metrics in Tab.~\ref{tab:chi2_per_axion}. Posterior distributions for $f_{\text{ax}}^i$ and $H_0$, with and without the H0DN prior, are shown in Fig.~\ref{fig:123npsd}, while the PL of $H_0$ computed from the PD dataset combination is shown in Fig.~\ref{fig:profile},. The total EDE fractional contribution as a function of redshift is shown in Fig.~\ref{fig:energy_injection}. Finally, in Fig.~\ref{fig:chi2perexp} we show the change in best-fit $\Delta\chi^2$ (total and per-likelihood) relative to $\Lambda$CDM as a function of the number of EDE fields, $N_{\text{ax}}$.
\begin{table}[h]
    \centering
    \begin{tabular}{|c? cc | c|}
    \hline
    & \multicolumn{2}{c|}{Credible intervals} & Confidence intervals \\
    \cline{2-3} \cline{4-4}
    Model & with H0DN & no H0DN & Bestfit $^+_-68\%$ $(95\%)$ \\
    \hline \hline

    $\Lambda$CDM & --- & $68.10\pm0.28$ & $68.10_{-0.40}^{+0.40}\,(_{-0.80}^{+0.80})$ \\[4pt]
    1 axion & $72.16^{+0.67}_{-0.68}$ & $69.53^{+0.71}_{-1.20}$ & $69.89_{-1.03}^{+1.19}\,(_{-1.99}^{+2.21})$ \\[4pt]
    2 axions & $72.65^{+0.73}_{-0.75}$ & $70.46^{+0.96}_{-1.27}$ & $70.59_{-1.24}^{+1.66}\,(_{-2.42}^{+3.36})$ \\[4pt]
    3 axions & $72.91^{+0.68}_{-0.73}$ & $70.93^{+1.06}_{-1.47}$ & $70.93_{-1.49}^{+1.67}\,(_{-2.73}^{+3.55})$ \\
    \hline    \hline

    \end{tabular}
    \caption{$H_0$ constraints (in km/s/Mpc) from the PD and PDH0 combinations. The first two columns show means and $68\%$ Bayesian credible intervals with and without the H0DN prior. The third column shows frequentist confidence intervals from profile likelihoods on PD at $68\%$ ($95\%$).}
    \label{tab:h0_summary}
\end{table}

The comparison between $\Lambda$CDM and the $N_{\rm ax}=1,2,3$ cases (see Tab.~\ref{tab:h0_summary}) shows that axions that are dynamically important prior to recombination boost the mean value of $H_0$, and do not just reduce the tension by increasing the uncertainty (see also Fig.~\ref{fig:123npsd}). In particular, without the H0DN prior, the posterior mean increases by $\sim$ 2.4 km/sec/Mpc with the inclusion of 2 axions, and $\sim 2.8$km/sec/Mpc when adding a third one.

\begin{table}[!htb]
%\footnotesize
\begin{tabular}{|c?c|c|c|}
\hline
Model & $Q_{\text{DMAP}}$ & $\Delta$AIC(PD) & $\Delta\text{AIC(PDH0)}$  \\
\hline\hline
$\Lambda$CDM & $6.3\sigma$ & $-0.86$ & $+32.52$\\ \hline
1 axion  & $2.6\sigma$ & \text{ref.\ model} & \text{ref.\ model} \\ \hline
2 axions  & $1.5\sigma$& $+2.96$ & $-1.70$ \\ \hline
3 axions & $1.3\sigma$ & $+6.92$ & $+1.76$ \\ \hline
\hline
\end{tabular}
\caption{Tension and model comparison metrics. $Q_{\text{DMAP}}$ measures the residual tension with the H0DN determination. The $\Delta\text{AIC}$ values are computed with respect to the 1-axion model, so $\Delta\text{AIC}<0$ indicates that the model in question is favored against the 1-axion model by the given combination.}
\label{tab:chi2_per_axion}
\end{table}

To check the role of prior volume effects, the previous numbers should be compared to the PL of $H_0$, computed from the H0DN-free PD combination and reported in Fig.~\ref{fig:profile}. Using the Neyman construction~\cite{Neyman:1937uhy}, we find the $68\%$ ($95\%$) confidence intervals (CIs) reported in Tab.~\ref{tab:h0_summary}. We find that prior volume effects are small for the 2- and 3-axion models, and are only noticeable in the smaller error bar of the credible intervals compared to the CIs.~\footnote{We checked that prior volume effect can become significant for higher number of fields, with e.g.\ the $N_{\text{ax}}=9$ case yielding Bayesian posteriors in perfect agreement with the H0DN determination \textit{even without an H0DN prior}.} Moreover, we see that, while the PL quickly degrades above $H_0\sim 72\,\text{km/s/Mpc}$ in the standard 1-axion EDE model, having two or more axions allows us to reach $H_0$ values in agreement with the H0DN determination without significantly degrading the fit to the H0DN-free PD combination.  Importantly, our PL analysis demonstrates that the fit remains good for larger values of $H_0$ than those suggested by Bayesian credible intervals. The residual tension is reduced to the $2.6\sigma$, $1.5\sigma$, and $1.3\sigma$ levels for the $N_{\text{ax}}=1,2,3$ models respectively (see Tab.~\ref{tab:chi2_per_axion}).

Since each extra field carries two additional parameters ($z_c^i$ and $f_{\text{ax}}^i$), the best-fit $\chi^2$ should decrease by $\vert \Delta\chi^2 \vert \geq 4$ per axion for the AIC to favor the more complex model. From Tab.~\ref{tab:chi2_per_axion} we see that this is achieved by the 2-axion model, which is favored over the 1-axion one by the AIC metric for the PDH0 combination. For $N_{\text{ax}} \geq 2$ the best-fit $\chi^2$ quickly flattens, so the (marginal) improvement in fit is insufficient to justify additional fields beyond the second. This result is non-trivial, as it demonstrates that one can not get an arbitrarily good fit by increasing the EDE field content, and suggests that our multi-field EDE framework is predictive: it could be confirmed or excluded with future higher precision data~\cite{SimonsObservatory:2018koc,SimonsObservatory:2019qwx}.

In Fig.~\ref{fig:energy_injection} we show reconstructed posteriors for the total EDE fraction, $f_{\text{EDE}}(z)\equiv\sum_i \rho_{\text{ax}}^i(z)/\rho_{\text{tot}}(z)$, where $\rho_{\text{ax}}^i(z)$ is the energy density of the $i$th EDE field. In agreement with earlier results~\cite{Poulin:2023lkg}, we see that the 1-axion case requires fairly large values for the critical fraction, $f_{\text{ax}}^1 \sim 0.125$, around matter-radiation equality [$\log_{10}(z_c^1) \sim 3.6$]. Introducing an additional axion spreads out the effects of EDE over a broader redshift range compared to the 1-axion case. The first axion now peaks at lower redshifts ($\log_{10}(z_c^1) \sim 3.4$), while the second does so at higher redshifts ($\log_{10}(z_c^2) \sim 3.9$). The heights of both peaks are comparable and lower compared to the 1-axion case, i.e.\ $f_{\text{ax}}^1 \sim f_{\text{ax}}^2 \sim 0.08$, so the energy injection is smoother overall. The effect of the third axion is not visible, confirming our earlier results.
 
We show the per-likelihood $\Delta\chi^2_{\min}$ for the different models in Fig.~\ref{fig:chi2perexp}. We note that, as expected, the first axion dramatically improves the fit to the H0DN determination, moderately improves the fit to Planck low-$\ell$ data, but worsens the fit to NPIPE high-$\ell$ data as observed previously~\cite{Efstathiou:2023fbn}. The second axion plays a key role in improving the fit to NPIPE high-$\ell$ data, while only marginally altering the fit to the other datasets, including H0DN. We find in particular (see Tab.~\ref{tab:PD_LCDM}) that the damping scale $r_d^\star$ decreases by about $1\sigma$ in the 2-axion and 3-axion models compared to the 1-axion one, as a consequence of more extended energy injection (compare with Fig.~2 of Ref.~\cite{Knox:2019rjx}). We find that it is the fit to the multipoles around $\ell\sim 800-1200$ that are most improved by the inclusion of the second axion (see Fig.~\ref{fig:chi2perbin}).
 
\textbf{\textit{Conclusions.}} --- We have shown that extending the simplest axion-like EDE model to include additional axion-like fields can significantly improve EDE's ability to address the Hubble tension, in light of the latest Planck NPIPE data, which strongly disfavor the 1-field scenario~\cite{Efstathiou:2023fbn}.  In a nutshell, we find that within a 2-axion EDE model the residual tension with the latest H0DN determination is reduced to barely $1.5\sigma$, compatible with a statistical fluctuation, and the model is favored over the 1-axion one. Including additional axions does not further improve the fit. This demonstrates that the difficulties of EDE in alleviating the Hubble tension when adopting NPIPE data reflect a limitation of the simplest single-field model, not of the framework itself.

Introducing a second axion leads to a smoother energy injection in the pre-recombination era, that improves the fit to Planck, with broader peaks than in the single-field case, see Fig.~\ref{fig:energy_injection}. We find that, while EDE cosmologies are known to predict larger values of $S_8$ compared to $\Lambda$CDM~\cite{Hill:2020osr,Vagnozzi:2021gjh,Vagnozzi:2023nrq,Poulin:2024ken,Pedrotti:2024kpn}, the ``$S_8$ discrepancy''~\cite{DiValentino:2020vvd,Pantos:2026koc} is not worsened by the addition of the second axion, remaining at the $\sim 2\sigma$ level when compared with the latest DES-Y6 results~\cite{DES:2026zjp}. As $\omega_b$ increases to maintain a good fit to the damping tail \cite{smith:2023oop,Giovanetti:2026aku}, the tension with the PRIMAT determination \cite{Pitrou:2019nub,Pitrou:2020etk} grows from 2.0$\sigma$ for $\Lambda$CDM to 2.8$\sigma$ for 2-field EDE (and remains in $1.3\sigma$ agreement with the PArthENoPE estimate \cite{Consiglio:2017pot}).

Interpreted more broadly, our results indicate that the Hubble tension may point to a complex pre-recombination expansion history, not captured by single-field EDE models. Extending our framework over a wider redshift range, confronting it with SPT-3G \cite{SPT-3G:2025bzu}, ACT DR6 \cite{AtacamaCosmologyTelescope:2025blo}, and large-scale structure data~\cite{DESI:2024hhd}, and exploring model-independent reconstructions of the pre-recombination expansion history~\cite{Moss:2021obd,Meiers:2023gft,An:2023buh,Kou:2024rvn}, are natural avenues for future work.

\textbf{\textit{Acknowledgments.}} --- V.P.\ acknowledges the European Union's Horizon Europe research and innovation programme under the Marie Sk\l odowska-Curie Staff Exchange grant agreement no.\ 101086085 -- ASYMMETRY. This work received funding support from the European Research Council (ERC) under the European Union's HORIZON-ERC-2022 (grant agreement no.\ 101076865). S.V.\ acknowledges support from the University of Trento and the Provincia Autonoma di Trento (PAT, Autonomous Province of Trento) through the UniTrento Internal Call for Research 2023 grant ``Searching for Dark Energy off the beaten track'' (DARKTRACK, grant agreement no.\ E63C22000500003), and from the Istituto Nazionale di Fisica Nucleare (INFN) through the Commissione Scientifica Nazionale 4 (CSN4) Iniziativa Specifica ``Quantum Fields in Gravity, Cosmology and Black Holes'' (FLAG). L.K.\ was supported in part by the Department of Energy (DOE) Office of Science award DESC0009999 and by the Michael and Ester Vaida Endowed Chair in Cosmology and Astrophysics. This publication is based upon work from the COST Action CA21136 ``Addressing observational tensions in cosmology with systematics and fundamental physics'' (CosmoVerse), supported by COST (European Cooperation in Science and Technology). The results obtained in this paper were computed through resources from the Universe and Particles Laboratory of Montpellier (LUPM). We thank LUPM for providing the technical support, computing and storage facilities.

\clearpage
\setcounter{figure}{0}
\renewcommand{\thefigure}{S\arabic{figure}}
\renewcommand{\theHfigure}{S\arabic{figure}}
\setcounter{table}{0}
\renewcommand{\thetable}{S\arabic{table}}
\renewcommand{\theHtable}{S\arabic{table}}
\setcounter{equation}{0}
\renewcommand{\theequation}{S\arabic{equation}}
\renewcommand{\theHequation}{S\arabic{equation}}
\section*{Supplementary Material}

\input{appendix}

\clearpage
\bibliography{multi_axion}

\end{document}

%% file: appendix.tex
\subsection*{\normalsize A. Full MCMC constraints}

In Tables~\ref{tab:PDH0_axion} and~\ref{tab:PDH0_LCDM} we report confidence intervals for the EDE parameters and the standard cosmological parameters respectively, adopting the PDH0 combination (Planck PR4+DESI DR2+PantheonPlus+H0DN). Tab.~\ref{tab:PD_LCDM} presents the corresponding cosmological parameter constraints for the PD combination (without the H0DN prior). In Tables~\ref{tab:PDH0_LCDM} and~\ref{tab:PD_LCDM}, we show confidence intervals obtained both within $\Lambda$CDM and the $N_{\text{ax}}=1,2,3$ models.

\begin{figure*}
\centering
\includegraphics[width=\linewidth]{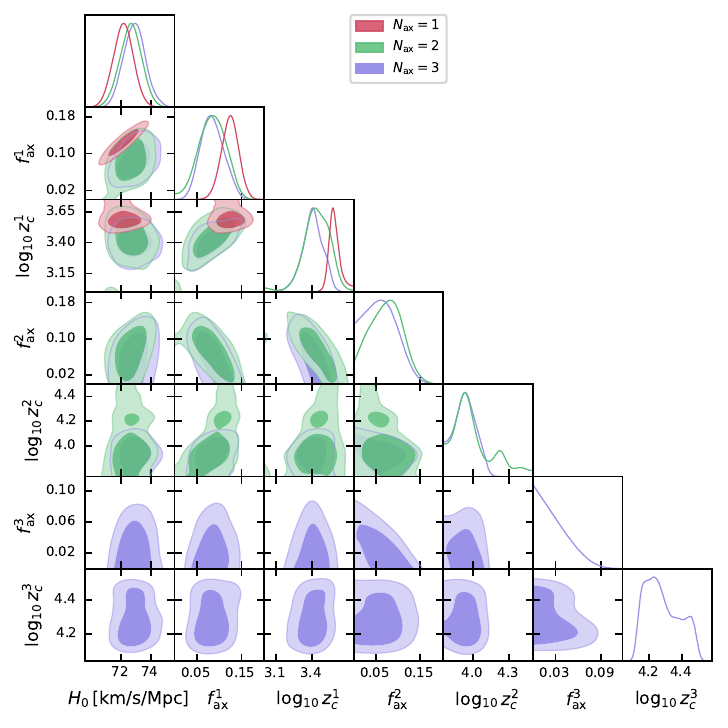}
\caption{Marginalized posteriors for the EDE parameters within the $N_{\text{ax}}=1,2,3$ models, in light of the PDH0 combination.}
\label{fig:edeparameterspdh0}
\end{figure*}

\begin{figure*}
\centering
\includegraphics[width=\linewidth]{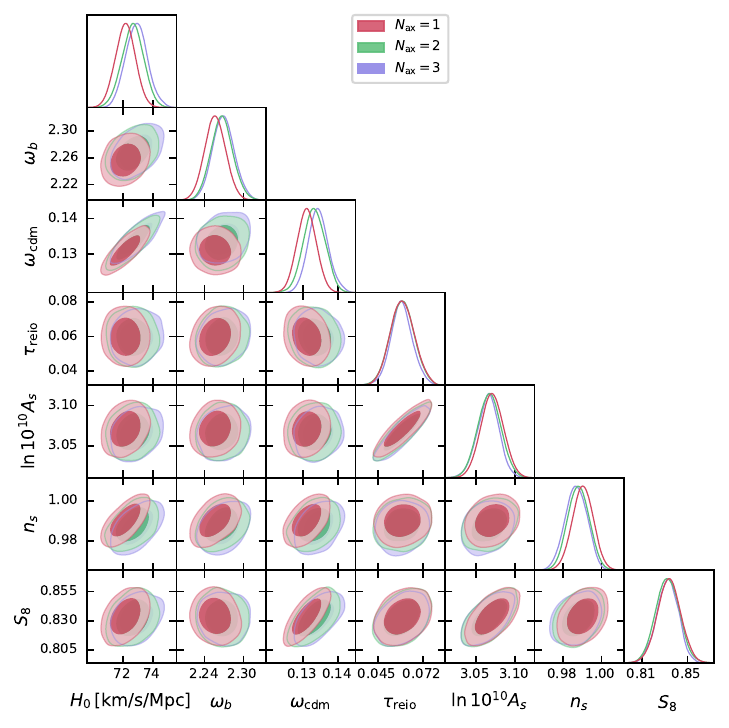}
\caption{As in Fig.~\ref{fig:edeparameterspdh0}, but for the standard cosmological parameters.}
\label{fig:parameterspdh0}
\end{figure*}

\begin{figure*}
\centering
\includegraphics[width=\linewidth]{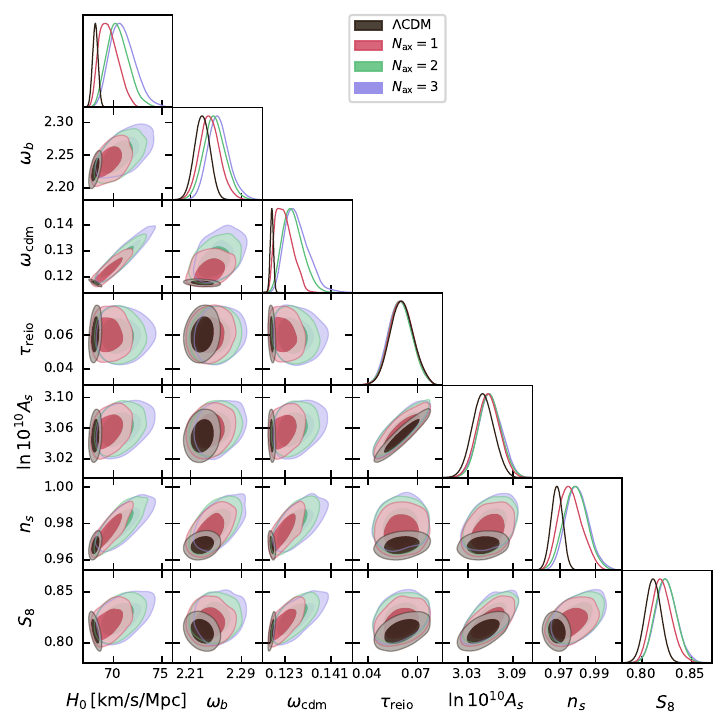}
\caption{As in Fig.~\ref{fig:parameterspdh0}, but in light of the PD combination.}
\label{fig:parameterspd}
\end{figure*}

The marginalized posteriors for the EDE parameters [$\log_{10}a_c^i$ and $f_{\mathrm{ax},i}$, where $a_c^i=(1+z_c^i)^{-1}$ is the critical scale factor] are shown in Fig.~\ref{fig:edeparameterspdh0} for the PDH0 combination. The corresponding posteriors for the standard cosmological parameters are shown in Fig.~\ref{fig:parameterspdh0}. Finally, Fig.~\ref{fig:parameterspd} shows the posteriors for the standard cosmological parameters obtained from the PD combination, comparing $\Lambda$CDM with the $N_{\text{ax}}=1,2,3$ models.

\begin{table*}
\begin{tabular}{c c cc cc}
\hline
 &  & \multicolumn{2}{c}{PDH0} & \multicolumn{2}{c}{PD} \\
Model & $i$ & $\log_{10}a_c^i$ & $f_{\rm ax,i}$ & $\log_{10}a_c^i$ & $f_{\rm ax,i}$ \\
\hline
\multirow{1}{*}{1 axion} & 1 & $-3.600^{+0.065}_{-0.028}$ & $0.124^{+0.019}_{-0.019}$ & $-3.683^{+0.210}_{-0.145}$ & $0.050^{+0.020}_{-0.043}$ \\ \hline
\multirow{2}{*}{2 axions} & 1 & $-3.440^{+0.091}_{-0.122}$ & $0.085^{+0.032}_{-0.030}$ & $-3.473^{+0.117}_{-0.137}$ & $0.049^{+0.019}_{-0.042}$ \\
 & 2 & $-4.000^{+0.217}_{-0.052}$ & $0.073^{+0.038}_{-0.038}$ & $-4.017^{+0.260}_{-0.225}$ & $<0.048$ \\
 \hline
\multirow{3}{*}{3 axions} & 1 & $-3.407^{+0.082}_{-0.091}$ & $0.085^{+0.024}_{-0.029}$ & $-3.444^{+0.114}_{-0.129}$ & $0.052^{+0.022}_{-0.041}$ \\
 & 2 & $-3.932^{+0.099}_{-0.098}$ & $0.062^{+0.029}_{-0.048}$ & $-3.910^{+0.139}_{-0.066}$ & $0.037^{+0.010}_{-0.037}$ \\
 & 3 & $-4.284^{+0.148}_{-0.170}$ & $<0.035$ & $-4.295^{+0.146}_{-0.078}$ & $<0.022$ \\ \hline
\end{tabular}
\caption{Means and $68\%$ Bayesian credible intervals for the EDE parameters $\log_{10}a_c^i$ and $f_{\mathrm{ax},i}$ in light of the PDH0 and PD combinations.}
\label{tab:PDH0_axion}
\end{table*}

\begin{table*}
\begin{tabular}{c cccc}
\multicolumn{5}{c}{PDH0}\\
\hline
Parameter & $\Lambda$CDM & 1 axion & 2 axions & 3 axions \\
\hline
$H_0$ & $68.661^{+0.255}_{-0.254}$ & $72.160^{+0.668}_{-0.681}$ & $72.645^{+0.731}_{-0.748}$ & $72.911^{+0.679}_{-0.728}$ \\
$\omega_b$ & $2.242^{+0.011}_{-0.011}$ & $2.257^{+0.016}_{-0.016}$ & $2.267^{+0.017}_{-0.017}$ & $2.269^{+0.016}_{-0.017}$ \\
$\omega_\text{cdm}$ & $0.117^{+0.001}_{-0.001}$ & $0.131^{+0.003}_{-0.003}$ & $0.133^{+0.003}_{-0.003}$ & $0.134^{+0.003}_{-0.003}$ \\
$\tau_\text{reio}$ & $0.064^{+0.007}_{-0.007}$ & $0.060^{+0.007}_{-0.007}$ & $0.060^{+0.007}_{-0.007}$ & $0.059^{+0.006}_{-0.006}$ \\
$\ln 10^{10}A_s$ & $3.056^{+0.015}_{-0.014}$ & $3.071^{+0.014}_{-0.014}$ & $3.067^{+0.015}_{-0.015}$ & $3.066^{+0.014}_{-0.014}$ \\
$n_s$ & $0.971^{+0.003}_{-0.003}$ & $0.990^{+0.005}_{-0.005}$ & $0.988^{+0.006}_{-0.006}$ & $0.987^{+0.006}_{-0.006}$ \\
$S_8$ & $0.801^{+0.008}_{-0.008}$ & $0.834^{+0.010}_{-0.010}$ & $0.832^{+0.010}_{-0.010}$ & $0.832^{+0.010}_{-0.009}$ \\
$r_s^\star$ & $145.759^{+0.176}_{-0.176}$ & $138.624^{+1.246}_{-1.232}$ & $137.597^{+1.378}_{-1.407}$ & $137.048^{+1.255}_{-1.231}$ \\
$r_d^\star$ & $45.057^{+0.081}_{-0.080}$ & $43.246^{+0.321}_{-0.324}$ & $42.927^{+0.366}_{-0.355}$ & $42.809^{+0.314}_{-0.323}$ \\ \hline
\end{tabular}
\caption{As in Tab.~\ref{tab:PDH0_axion}, but for the standard $\Lambda$CDM parameters. We also report the corresponding constraints on the derived parameters $S_8=\sigma_8(\Omega_m/0.3)^{0.5}$, the sound horizon at recombination $r_s^\star$, and the damping scale at recombination $r_d^\star$.}
\label{tab:PDH0_LCDM}
\end{table*}

\begin{table*}
\begin{tabular}{c cccc}
\multicolumn{5}{c}{PD}\\
\hline
Parameter & $\Lambda$CDM & 1 axion & 2 axions & 3 axions \\
\hline
$H_0$ & $68.103^{+0.278}_{-0.278}$ & $69.525^{+0.715}_{-1.201}$ & $70.459^{+0.961}_{-1.273}$ & $70.932^{+1.061}_{-1.471}$ \\
$\omega_b$ & $2.228^{+0.012}_{-0.012}$ & $2.240^{+0.015}_{-0.017}$ & $2.248^{+0.016}_{-0.018}$ & $2.254^{+0.015}_{-0.018}$ \\
$\omega_\text{cdm}$ & $0.118^{+0.001}_{-0.001}$ & $0.123^{+0.002}_{-0.004}$ & $0.126^{+0.003}_{-0.005}$ & $0.128^{+0.003}_{-0.006}$ \\
$\tau_\text{reio}$ & $0.060^{+0.007}_{-0.007}$ & $0.060^{+0.007}_{-0.007}$ & $0.060^{+0.007}_{-0.007}$ & $0.060^{+0.007}_{-0.008}$ \\
$\ln 10^{10}A_s$ & $3.050^{+0.014}_{-0.014}$ & $3.057^{+0.015}_{-0.015}$ & $3.059^{+0.014}_{-0.015}$ & $3.059^{+0.015}_{-0.016}$ \\
$n_s$ & $0.968^{+0.003}_{-0.003}$ & $0.976^{+0.005}_{-0.007}$ & $0.979^{+0.007}_{-0.007}$ & $0.980^{+0.006}_{-0.007}$ \\
$S_8$ & $0.811^{+0.008}_{-0.008}$ & $0.820^{+0.009}_{-0.011}$ & $0.824^{+0.010}_{-0.011}$ & $0.825^{+0.010}_{-0.011}$ \\
$r_s^\star$ & $145.555^{+0.179}_{-0.179}$ & $142.950^{+2.274}_{-1.098}$ & $141.208^{+2.239}_{-1.701}$ & $140.411^{+2.609}_{-1.799}$ \\
$r_d^\star$ & $45.054^{+0.081}_{-0.080}$ & $44.390^{+0.539}_{-0.324}$ & $43.931^{+0.591}_{-0.450}$ & $43.682^{+0.701}_{-0.448}$ \\ \hline
\end{tabular}
\caption{As in Tab.~\ref{tab:PDH0_LCDM}, but in light of the PD combination.}
\label{tab:PD_LCDM}
\end{table*}

\subsection*{\normalsize B. $\chi^2$-per-bin}

\begin{figure}
\centering
\includegraphics[width=\linewidth]{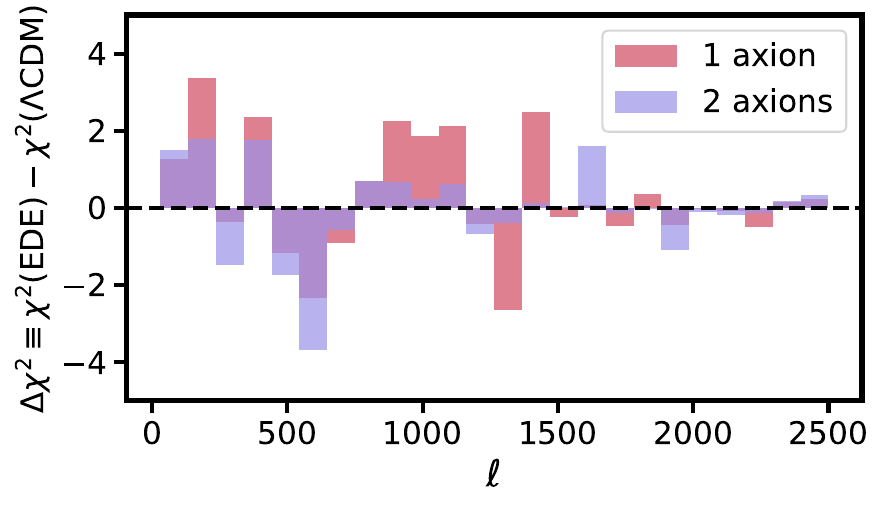}
\caption{Contribution to the total $\Delta\chi^2_{\rm min}=\chi^2({\rm EDE})-\chi^2(\Lambda{\rm CDM})$ for each CMB multipole bin, assuming a width $\Delta \ell=100$. We compare the best-fit 2-axion model fit to PDH0 ($H_0=72.97\,\text{km/s/Mpc}$) and a nearby 1-axion profile likelihood point ($H_0=72.99\,\text{km/s/Mpc}$), against the best-fit $\Lambda$CDM model fit to the PD combination ($H_0=68.10\,\text{km/s/Mpc}$).}
\label{fig:chi2perbin}
\end{figure}

In Fig.~\ref{fig:chi2perbin} we plot the contribution to the total $\Delta\chi^2_{\rm min}=\chi^2({\rm EDE})-\chi^2(\Lambda{\rm CDM})$ for each CMB multipole bin, assuming a width $\Delta \ell=100$. We compare the best-fit 2-axion model fit to PDH0 ($H_0 = 72.97\,\text{km/s/Mpc}$) and a nearby 1-axion profile likelihood point ($H_0=72.99\,\text{km/s/Mpc}$), against the best-fit $\Lambda$CDM model fit to the PD combination ($H_0=68.10\,\text{km/s/Mpc}$). One can see that the multipoles around $\ell\sim1000$ are those for which the fit is most improved within the 2-axion model compared to the 1-axion model.

\subsection*{\normalsize C. $\chi^2$-per-axion}

In Tab.~\ref{tab:chi2perexperiment} we provide the best-fit $\chi^2_{\min}$ per likelihood, for all models and dataset combinations considered.

\begin{table*}[h!]
\centering
\begin{tabular}{|l|cc|cc|cc|cc|}
\hline
&\multicolumn{2}{c}{$\Lambda$CDM} & \multicolumn{2}{c}{$N_{\text{ax}}=1$} & \multicolumn{2}{c}{$N_{\text{ax}}=2$} & \multicolumn{2}{c|}{$N_{\text{ax}}=3$}\\\hline
Dataset & PDH0 & PD & PDH0 & PD & PDH0 & PD & PDH0 & PD\\\hline
NPIPE high-$\ell$ TTTEEE & 11244.98 & 11241.12 & 11246.11 & 11240.04 & 11241.76 & 11239.17 & 11241.57 & 11239.27 \\
Planck 18 low-$\ell$ EE & 398.34 & 397.27 & 396.83 & 396.88 & 396.64 & 396.88 & 396.80 & 396.91 \\
Planck 18 low-$\ell$ TT & 22.42 & 22.79 & 20.97 & 21.91 & 20.91 & 21.63 & 21.14 & 21.45 \\
PantheonPlus & 1414.05 & 1412.53 & 1414.09 & 1412.99 & 1413.91 & 1413.24 & 1414.03 & 1413.29 \\
DESI & 10.32 & 13.23 & 10.47 & 12.07 & 10.99 & 11.69 & 10.87 & 11.70 \\
H0DN & 36.03 &  & 2.09 &  & 0.42 &  & 0.15 &  \\\hline
total $\chi^2_\text{min}$ & 13136.98 & 13097.01 & 13100.47 & 13093.73 & 13094.75 & 13092.54 & 13094.41 & 13092.51 \\ \hline$\chi^2_\text{min}-\chi^2_\text{min}(N_\text{ax}=1)$ &  36.51 & 3.28 & \multicolumn{2}{c|}{ref.\ model}  & $-5.71$ & $-1.18$ & $-6.06$ & $-1.22$\\ \hline
$Q_{\rm DMAP}$ & \multicolumn{2}{c|}{6.3$\sigma$} & \multicolumn{2}{c|}{2.6$\sigma$}& \multicolumn{2}{c|}{1.5$\sigma$}&\multicolumn{2}{c|}{1.3$\sigma$} \\ \hline
\end{tabular}
\caption{Best-fit per-likelihood and total $\chi^2$ for the $N_{\text{ax}}=1,2,3$ models, in light of the PDH0 and PD combinations.}
\label{tab:chi2perexperiment}
\end{table*}

%% file: multi_axion.bib
@article{H0DN:2025lyy,
    author = "Casertano, Stefano and others",
    collaboration = "H0DN",
    title = "{The Local Distance Network: a community consensus report on the measurement of the Hubble constant at 1{\%} precision}",
    eprint = "2510.23823",
    archivePrefix = "arXiv",
    primaryClass = "astro-ph.CO",
    month = "10",
    year = "2025"
}

@article{SPT-3G:2025vyw,
    author = "Khalife, A. R. and others",
    collaboration = "SPT-3G",
    title = "{SPT-3G D1: Axion Early Dark Energy with CMB experiments and DESI}",
    eprint = "2507.23355",
    archivePrefix = "arXiv",
    primaryClass = "astro-ph.CO",
    reportNumber = "FERMILAB-PUB-25-0610-PPD",
    month = "7",
    year = "2025"
}

@article{Consiglio:2017pot,
    author = "Consiglio, R. and de Salas, P. F. and Mangano, G. and Miele, G. and Pastor, S. and Pisanti, O.",
    title = "{PArthENoPE reloaded}",
    eprint = "1712.04378",
    archivePrefix = "arXiv",
    primaryClass = "astro-ph.CO",
    doi = "10.1016/j.cpc.2018.06.022",
    journal = "Comput. Phys. Commun.",
    volume = "233",
    pages = "237--242",
    year = "2018"
}

@article{Pitrou:2019nub,
    author = "Pitrou, Cyril and Coc, Alain and Uzan, Jean-Philippe and Vangioni, Elisabeth",
    editor = "Kawabata, T. and others",
    title = "{Precision Big Bang Nucleosynthesis with the New Code PRIMAT}",
    eprint = "1909.12046",
    archivePrefix = "arXiv",
    primaryClass = "astro-ph.CO",
    doi = "10.7566/JPSCP.31.011034",
    journal = "JPS Conf. Proc.",
    volume = "31",
    pages = "011034",
    year = "2020"
}

@article{Pitrou:2020etk,
    author = "Pitrou, Cyril and Coc, Alain and Uzan, Jean-Philippe and Vangioni, Elisabeth",
    title = "{A new tension in the cosmological model from primordial deuterium?}",
    eprint = "2011.11320",
    archivePrefix = "arXiv",
    primaryClass = "astro-ph.CO",
    doi = "10.1093/mnras/stab135",
    journal = "Mon. Not. Roy. Astron. Soc.",
    volume = "502",
    number = "2",
    pages = "2474--2481",
    year = "2021"
}

@article{McDonough:2023qcu,
    author = "McDonough, Evan and Hill, J. Colin and Ivanov, Mikhail M. and La Posta, Adrien and Toomey, Michael W.",
    title = "{Observational constraints on early dark energy}",
    eprint = "2310.19899",
    archivePrefix = "arXiv",
    primaryClass = "astro-ph.CO",
    reportNumber = "MIT-CTP/5618",
    doi = "10.1142/S0218271824300039",
    journal = "Int. J. Mod. Phys. D",
    volume = "33",
    number = "11",
    pages = "2430003",
    year = "2024"
}

@article{Giovanetti:2026aku,
    author = "Giovanetti, Cara",
    title = "{A generic $ω_b$ tension in early-time solutions to the Hubble tension}",
    eprint = "2604.05095",
    archivePrefix = "arXiv",
    primaryClass = "astro-ph.CO",
    month = "4",
    year = "2026"
}

@article{Smith:2023oop,
    author = "Smith, Tristan L. and Poulin, Vivian",
    title = "{Current small-scale CMB constraints to axionlike early dark energy}",
    eprint = "2309.03265",
    archivePrefix = "arXiv",
    primaryClass = "astro-ph.CO",
    doi = "10.1103/PhysRevD.109.103506",
    journal = "Phys. Rev. D",
    volume = "109",
    number = "10",
    pages = "103506",
    year = "2024"
}

@article{Moss:2021obd,
    author = "Moss, Adam and Copeland, Edmund and Bamford, Steven and Clarke, Thomas",
    title = "{A model-independent reconstruction of dark energy to very high redshift}",
    eprint = "2109.14848",
    archivePrefix = "arXiv",
    primaryClass = "astro-ph.CO",
    month = "9",
    year = "2021"
}

@article{Meiers:2023gft,
    author = {Meiers, Michael and Knox, Lloyd and Sch{\"o}neberg, Nils},
    title = "{Exploration of the prerecombination universe with a high-dimensional model of an additional dark fluid}",
    eprint = "2307.09522",
    archivePrefix = "arXiv",
    primaryClass = "astro-ph.CO",
    doi = "10.1103/PhysRevD.108.103527",
    journal = "Phys. Rev. D",
    volume = "108",
    number = "10",
    pages = "103527",
    year = "2023"
}

@article{Kou:2024rvn,
    author = {Kou, Rapha{\"e}l and Lewis, Antony},
    title = "{A flexible parameterization to test early physics solutions to the Hubble tension with future CMB data}",
    eprint = "2410.16185",
    archivePrefix = "arXiv",
    primaryClass = "astro-ph.CO",
    doi = "10.1088/1475-7516/2025/01/033",
    journal = "JCAP",
    volume = "01",
    pages = "033",
    year = "2025"
}

@article{Lesgourgues:2011re,
    author = "Lesgourgues, Julien",
    title = "{The Cosmic Linear Anisotropy Solving System (CLASS) I: Overview}",
    eprint = "1104.2932",
    archivePrefix = "arXiv",
    primaryClass = "astro-ph.IM",
    month = "4",
    year = "2011"
}

@article{Simon:2023hlp,
    author = "Simon, Th{\'e}o",
    title = "{Can acoustic and axionlike early dark energy still resolve the Hubble tension?}",
    eprint = "2310.16800",
    archivePrefix = "arXiv",
    primaryClass = "astro-ph.CO",
    doi = "10.1103/PhysRevD.110.023528",
    journal = "Phys. Rev. D",
    volume = "110",
    number = "2",
    pages = "023528",
    year = "2024"
}

@article{Brinckmann:2018cvx,
    author = "Brinckmann, Thejs and Lesgourgues, Julien",
    title = "{MontePython 3: boosted MCMC sampler and other features}",
    eprint = "1804.07261",
    archivePrefix = "arXiv",
    primaryClass = "astro-ph.CO",
    reportNumber = "TTK-18-15",
    doi = "10.1016/j.dark.2018.100260",
    journal = "Phys. Dark Univ.",
    volume = "24",
    pages = "100260",
    year = "2019"
}

@article{Audren:2012wb,
    author = "Audren, Benjamin and Lesgourgues, Julien and Benabed, Karim and Prunet, Simon",
    title = "{Conservative Constraints on Early Cosmology: an illustration of the Monte Python cosmological parameter inference code}",
    eprint = "1210.7183",
    archivePrefix = "arXiv",
    primaryClass = "astro-ph.CO",
    reportNumber = "CERN-PH-TH-2012-290, LAPTH-048-12",
    doi = "10.1088/1475-7516/2013/02/001",
    journal = "JCAP",
    volume = "02",
    pages = "001",
    year = "2013"
}

@article{Lewis:2019xzd,
    author = "Lewis, Antony",
    title = "{GetDist: a Python package for analysing Monte Carlo samples}",
    eprint = "1910.13970",
    archivePrefix = "arXiv",
    primaryClass = "astro-ph.IM",
    doi = "10.1088/1475-7516/2025/08/025",
    journal = "JCAP",
    volume = "08",
    pages = "025",
    year = "2025"
}

@article{Brout:2022vxf,
    author = "Brout, Dillon and others",
    title = "{The Pantheon+ Analysis: Cosmological Constraints}",
    eprint = "2202.04077",
    archivePrefix = "arXiv",
    primaryClass = "astro-ph.CO",
    doi = "10.3847/1538-4357/ac8e04",
    journal = "Astrophys. J.",
    volume = "938",
    number = "2",
    pages = "110",
    year = "2022"
}

@article{Smith:2020rxx,
    author = "Smith, Tristan L. and Poulin, Vivian and Bernal, Jos{\'e} Luis and Boddy, Kimberly K. and Kamionkowski, Marc and Murgia, Riccardo",
    title = "{Early dark energy is not excluded by current large-scale structure data}",
    eprint = "2009.10740",
    archivePrefix = "arXiv",
    primaryClass = "astro-ph.CO",
    doi = "10.1103/PhysRevD.103.123542",
    journal = "Phys. Rev. D",
    volume = "103",
    number = "12",
    pages = "123542",
    year = "2021"
}

@article{Reeves:2022aoi,
    author = "Reeves, Alexander and Herold, Laura and Vagnozzi, Sunny and Sherwin, Blake D. and Ferreira, Elisa G. M.",
    title = "{Restoring cosmological concordance with early dark energy and massive neutrinos?}",
    eprint = "2207.01501",
    archivePrefix = "arXiv",
    primaryClass = "astro-ph.CO",
    doi = "10.1093/mnras/stad317",
    journal = "Mon. Not. Roy. Astron. Soc.",
    volume = "520",
    number = "3",
    pages = "3688--3695",
    year = "2023"
}

@article{Herold:2024enb,
    author = "Herold, Laura and Ferreira, Elisa G. M. and Heinrich, Lukas",
    title = "{Profile likelihoods in cosmology: When, why, and how illustrated with {\ensuremath{\Lambda}}CDM, massive neutrinos, and dark energy}",
    eprint = "2408.07700",
    archivePrefix = "arXiv",
    primaryClass = "astro-ph.CO",
    doi = "10.1103/PhysRevD.111.083504",
    journal = "Phys. Rev. D",
    volume = "111",
    number = "8",
    pages = "083504",
    year = "2025"
}

@article{Herold:2021ksg,
    author = "Herold, Laura and Ferreira, Elisa G. M. and Komatsu, Eiichiro",
    title = "{New Constraint on Early Dark Energy from Planck and BOSS Data Using the Profile Likelihood}",
    eprint = "2112.12140",
    archivePrefix = "arXiv",
    primaryClass = "astro-ph.CO",
    doi = "10.3847/2041-8213/ac63a3",
    journal = "Astrophys. J. Lett.",
    volume = "929",
    number = "1",
    pages = "L16",
    year = "2022"
}

@article{Gomez-Valent:2022hkb,
    author = "G{\'o}mez-Valent, Adri{\`a}",
    title = "{Fast test to assess the impact of marginalization in Monte~Carlo analyses and its application to cosmology}",
    eprint = "2203.16285",
    archivePrefix = "arXiv",
    primaryClass = "astro-ph.CO",
    doi = "10.1103/PhysRevD.106.063506",
    journal = "Phys. Rev. D",
    volume = "106",
    number = "6",
    pages = "063506",
    year = "2022"
}

@article{Karwal:2024qpt,
    author = "Karwal, Tanvi and Patel, Yashvi and Bartlett, Alexa and Poulin, Vivian and Smith, Tristan L. and Pfeffer, Daniel N.",
    title = "{Procoli: Profiles of cosmological likelihoods}",
    eprint = "2401.14225",
    archivePrefix = "arXiv",
    primaryClass = "astro-ph.CO",
    month = "1",
    year = "2024"
}

@article{Schoneberg:2021qvd,
    author = {Sch{\"o}neberg, Nils and Franco Abell{\'a}n, Guillermo and P{\'e}rez S{\'a}nchez, Andrea and Witte, Samuel J. and Poulin, Vivian and Lesgourgues, Julien},
    title = "{The H0 Olympics: A fair ranking of proposed models}",
    eprint = "2107.10291",
    archivePrefix = "arXiv",
    primaryClass = "astro-ph.CO",
    doi = "10.1016/j.physrep.2022.07.001",
    journal = "Phys. Rept.",
    volume = "984",
    pages = "1--55",
    year = "2022"
}

@article{DES:2026zjp,
    author = "Sanchez-Cid, D. and others",
    collaboration = "DES",
    title = "{Dark Energy Survey Year 6 Results: Weak Lensing and Galaxy Clustering Cosmological Analysis Framework}",
    eprint = "2601.14859",
    archivePrefix = "arXiv",
    primaryClass = "astro-ph.CO",
    reportNumber = "FERMILAB-PUB-26-0035-PPD",
    month = "1",
    year = "2026"
}

@article{Raveri:2018wln,
    author = "Raveri, Marco and Hu, Wayne",
    title = "{Concordance and Discordance in Cosmology}",
    eprint = "1806.04649",
    archivePrefix = "arXiv",
    primaryClass = "astro-ph.CO",
    doi = "10.1103/PhysRevD.99.043506",
    journal = "Phys. Rev. D",
    volume = "99",
    number = "4",
    pages = "043506",
    year = "2019"
}

@article{Karwal:2016vyq,
    author = "Karwal, Tanvi and Kamionkowski, Marc",
    title = "{Dark energy at early times, the Hubble parameter, and the string axiverse}",
    eprint = "1608.01309",
    archivePrefix = "arXiv",
    primaryClass = "astro-ph.CO",
    doi = "10.1103/PhysRevD.94.103523",
    journal = "Phys. Rev. D",
    volume = "94",
    number = "10",
    pages = "103523",
    year = "2016"
}

@article{Poulin:2018cxd,
    author = "Poulin, Vivian and Smith, Tristan L. and Karwal, Tanvi and Kamionkowski, Marc",
    title = "{Early Dark Energy Can Resolve The Hubble Tension}",
    eprint = "1811.04083",
    archivePrefix = "arXiv",
    primaryClass = "astro-ph.CO",
    doi = "10.1103/PhysRevLett.122.221301",
    journal = "Phys. Rev. Lett.",
    volume = "122",
    number = "22",
    pages = "221301",
    year = "2019"
}

@article{Smith:2019ihp,
    author = "Smith, Tristan L. and Poulin, Vivian and Amin, Mustafa A.",
    title = "{Oscillating scalar fields and the Hubble tension: a resolution with novel signatures}",
    eprint = "1908.06995",
    archivePrefix = "arXiv",
    primaryClass = "astro-ph.CO",
    doi = "10.1103/PhysRevD.101.063523",
    journal = "Phys. Rev. D",
    volume = "101",
    number = "6",
    pages = "063523",
    year = "2020"
}

@article{Poulin:2025nfb,
    author = "Poulin, Vivian and Smith, Tristan L. and Calder{\'o}n, Rodrigo and Simon, Th{\'e}o",
    title = "{Impact of ACT DR6 and DESI DR2 for early dark energy and the Hubble tension}",
    eprint = "2505.08051",
    archivePrefix = "arXiv",
    primaryClass = "astro-ph.CO",
    doi = "10.1103/bx25-1g5d",
    journal = "Phys. Rev. D",
    volume = "113",
    number = "6",
    pages = "063519",
    year = "2026"
}

@article{Planck:2020olo,
    author = "Akrami, Y. and others",
    collaboration = "Planck",
    title = "{$Planck$ intermediate results. LVII. Joint Planck LFI and HFI data processing}",
    eprint = "2007.04997",
    archivePrefix = "arXiv",
    primaryClass = "astro-ph.CO",
    doi = "10.1051/0004-6361/202038073",
    journal = "Astron. Astrophys.",
    volume = "643",
    pages = "A42",
    year = "2020"
}

@article{Rosenberg:2022sdy,
    author = "Rosenberg, Erik and Gratton, Steven and Efstathiou, George",
    title = "{CMB power spectra and cosmological parameters from Planck PR4 with CamSpec}",
    eprint = "2205.10869",
    archivePrefix = "arXiv",
    primaryClass = "astro-ph.CO",
    doi = "10.1093/mnras/stac2744",
    journal = "Mon. Not. Roy. Astron. Soc.",
    volume = "517",
    number = "3",
    pages = "4620--4636",
    year = "2022"
}

@article{Efstathiou:2023fbn,
    author = "Efstathiou, George and Rosenberg, Erik and Poulin, Vivian",
    title = "{Improved Planck Constraints on Axionlike Early Dark Energy as a Resolution of the Hubble Tension}",
    eprint = "2311.00524",
    archivePrefix = "arXiv",
    primaryClass = "astro-ph.CO",
    doi = "10.1103/PhysRevLett.132.221002",
    journal = "Phys. Rev. Lett.",
    volume = "132",
    number = "22",
    pages = "221002",
    year = "2024"
}

@article{Arvanitaki:2009fg,
    author = "Arvanitaki, Asimina and Dimopoulos, Savas and Dubovsky, Sergei and Kaloper, Nemanja and March-Russell, John",
    title = "{String Axiverse}",
    eprint = "0905.4720",
    archivePrefix = "arXiv",
    primaryClass = "hep-th",
    doi = "10.1103/PhysRevD.81.123530",
    journal = "Phys. Rev. D",
    volume = "81",
    pages = "123530",
    year = "2010"
}

@article{SPT-3G:2025bzu,
    author = "Camphuis, E. and others",
    collaboration = "SPT-3G",
    title = "{SPT-3G D1: CMB temperature and polarization power spectra and cosmology from 2019 and 2020 observations of the SPT-3G Main field}",
    eprint = "2506.20707",
    archivePrefix = "arXiv",
    primaryClass = "astro-ph.CO",
    reportNumber = "FERMILAB-PUB-25-0144-PPD",
    month = "6",
    year = "2025"
}

@article{DESI:2025zgx,
    author = "Abdul Karim, M. and others",
    collaboration = "DESI",
    title = "{DESI DR2 results. II. Measurements of baryon acoustic oscillations and cosmological constraints}",
    eprint = "2503.14738",
    archivePrefix = "arXiv",
    primaryClass = "astro-ph.CO",
    reportNumber = "FERMILAB-PUB-25-0169-PPD",
    doi = "10.1103/tr6y-kpc6",
    journal = "Phys. Rev. D",
    volume = "112",
    number = "8",
    pages = "083515",
    year = "2025"
}

@article{Poulin:2023lkg,
    author = "Poulin, Vivian and Smith, Tristan L. and Karwal, Tanvi",
    title = "{The Ups and Downs of Early Dark Energy solutions to the Hubble tension: A review of models, hints and constraints circa 2023}",
    eprint = "2302.09032",
    archivePrefix = "arXiv",
    primaryClass = "astro-ph.CO",
    doi = "10.1016/j.dark.2023.101348",
    journal = "Phys. Dark Univ.",
    volume = "42",
    pages = "101348",
    year = "2023"
}

@article{Lee:2022gzh,
    author = {Lee, Nanoom and Ali-Ha{\"\i}moud, Yacine and Sch{\"o}neberg, Nils and Poulin, Vivian},
    title = "{What It Takes to Solve the Hubble Tension through Modifications of Cosmological Recombination}",
    eprint = "2212.04494",
    archivePrefix = "arXiv",
    primaryClass = "astro-ph.CO",
    doi = "10.1103/PhysRevLett.130.161003",
    journal = "Phys. Rev. Lett.",
    volume = "130",
    number = "16",
    pages = "161003",
    year = "2023"
}

@article{Vagnozzi:2021gjh,
    author = "Vagnozzi, Sunny",
    title = "{Consistency tests of {\ensuremath{\Lambda}}CDM from the early integrated Sachs-Wolfe effect: Implications for early-time new physics and the Hubble tension}",
    eprint = "2105.10425",
    archivePrefix = "arXiv",
    primaryClass = "astro-ph.CO",
    doi = "10.1103/PhysRevD.104.063524",
    journal = "Phys. Rev. D",
    volume = "104",
    number = "6",
    pages = "063524",
    year = "2021"
}

@article{Poulin:2024ken,
    author = "Poulin, Vivian and Smith, Tristan L. and Calder{\'o}n, Rodrigo and Simon, Th{\'e}o",
    title = "{Implications of the cosmic calibration tension beyond H0 and the synergy between early- and late-time new physics}",
    eprint = "2407.18292",
    archivePrefix = "arXiv",
    primaryClass = "astro-ph.CO",
    doi = "10.1103/PhysRevD.111.083552",
    journal = "Phys. Rev. D",
    volume = "111",
    number = "8",
    pages = "083552",
    year = "2025"
}

@article{Pedrotti:2024kpn,
    author = "Pedrotti, Davide and Jiang, Jun-Qian and Escamilla, Luis A. and da Costa, Simony Santos and Vagnozzi, Sunny",
    title = "{Multidimensionality of the Hubble tension: The roles of {\ensuremath{\Omega}}m and {\ensuremath{\omega}}c}",
    eprint = "2408.04530",
    archivePrefix = "arXiv",
    primaryClass = "astro-ph.CO",
    doi = "10.1103/PhysRevD.111.023506",
    journal = "Phys. Rev. D",
    volume = "111",
    number = "2",
    pages = "023506",
    year = "2025"
}

@article{Vagnozzi:2023nrq,
    author = "Vagnozzi, Sunny",
    title = "{Seven Hints That Early-Time New Physics Alone Is Not Sufficient to Solve the Hubble Tension}",
    eprint = "2308.16628",
    archivePrefix = "arXiv",
    primaryClass = "astro-ph.CO",
    doi = "10.3390/universe9090393",
    journal = "Universe",
    volume = "9",
    number = "9",
    pages = "393",
    year = "2023"
}

@article{Bernal:2016gxb,
    author = "Bernal, Jose Luis and Verde, Licia and Riess, Adam G.",
    title = "{The trouble with $H_0$}",
    eprint = "1607.05617",
    archivePrefix = "arXiv",
    primaryClass = "astro-ph.CO",
    doi = "10.1088/1475-7516/2016/10/019",
    journal = "JCAP",
    volume = "10",
    pages = "019",
    year = "2016"
}

@article{Lemos:2018smw,
    author = "Lemos, Pablo and Lee, Elizabeth and Efstathiou, George and Gratton, Steven",
    title = "{Model independent $H(z)$ reconstruction using the cosmic inverse distance ladder}",
    eprint = "1806.06781",
    archivePrefix = "arXiv",
    primaryClass = "astro-ph.CO",
    doi = "10.1093/mnras/sty3082",
    journal = "Mon. Not. Roy. Astron. Soc.",
    volume = "483",
    number = "4",
    pages = "4803--4810",
    year = "2019"
}

@article{Aylor:2018drw,
    author = "Aylor, Kevin and Joy, MacKenzie and Knox, Lloyd and Millea, Marius and Raghunathan, Srinivasan and Wu, W. L. Kimmy",
    title = "{Sounds Discordant: Classical Distance Ladder \& $\Lambda$CDM -based Determinations of the Cosmological Sound Horizon}",
    eprint = "1811.00537",
    archivePrefix = "arXiv",
    primaryClass = "astro-ph.CO",
    doi = "10.3847/1538-4357/ab0898",
    journal = "Astrophys. J.",
    volume = "874",
    number = "1",
    pages = "4",
    year = "2019"
}

@article{Schoneberg:2019wmt,
    author = {Sch\"oneberg, Nils and Lesgourgues, Julien and Hooper, Deanna C.},
    title = "{The BAO+BBN take on the Hubble tension}",
    eprint = "1907.11594",
    archivePrefix = "arXiv",
    primaryClass = "astro-ph.CO",
    reportNumber = "TTK-19-30",
    doi = "10.1088/1475-7516/2019/10/029",
    journal = "JCAP",
    volume = "10",
    pages = "029",
    year = "2019"
}

@article{Knox:2019rjx,
    author = "Knox, Lloyd and Millea, Marius",
    title = "{Hubble constant hunter\textquoteright{}s guide}",
    eprint = "1908.03663",
    archivePrefix = "arXiv",
    primaryClass = "astro-ph.CO",
    doi = "10.1103/PhysRevD.101.043533",
    journal = "Phys. Rev. D",
    volume = "101",
    number = "4",
    pages = "043533",
    year = "2020"
}

@article{Arendse:2019hev,
    author = "Arendse, Nikki and others",
    title = "{Cosmic dissonance: are new physics or systematics behind a short sound horizon?}",
    eprint = "1909.07986",
    archivePrefix = "arXiv",
    primaryClass = "astro-ph.CO",
    doi = "10.1051/0004-6361/201936720",
    journal = "Astron. Astrophys.",
    volume = "639",
    pages = "A57",
    year = "2020"
}

@article{Cai:2021weh,
    author = "Cai, Rong-Gen and Guo, Zong-Kuan and Wang, Shao-Jiang and Yu, Wang-Wei and Zhou, Yong",
    title = "{No-go guide for the Hubble tension: Late-time solutions}",
    eprint = "2107.13286",
    archivePrefix = "arXiv",
    primaryClass = "astro-ph.CO",
    doi = "10.1103/PhysRevD.105.L021301",
    journal = "Phys. Rev. D",
    volume = "105",
    number = "2",
    pages = "L021301",
    year = "2022"
}

@article{Keeley:2022ojz,
    author = "Keeley, Ryan E. and Shafieloo, Arman",
    title = "{Ruling Out New Physics at Low Redshift as a Solution to the H0 Tension}",
    eprint = "2206.08440",
    archivePrefix = "arXiv",
    primaryClass = "astro-ph.CO",
    doi = "10.1103/PhysRevLett.131.111002",
    journal = "Phys. Rev. Lett.",
    volume = "131",
    number = "11",
    pages = "111002",
    year = "2023"
}

@article{Jiang:2024xnu,
    author = "Jiang, Jun-Qian and Pedrotti, Davide and da Costa, Simony Santos and Vagnozzi, Sunny",
    title = "{Nonparametric late-time expansion history reconstruction and implications for the Hubble tension in light of recent DESI and type Ia supernovae data}",
    eprint = "2408.02365",
    archivePrefix = "arXiv",
    primaryClass = "astro-ph.CO",
    doi = "10.1103/PhysRevD.110.123519",
    journal = "Phys. Rev. D",
    volume = "110",
    number = "12",
    pages = "123519",
    year = "2024"
}

@article{Zhou:2025kws,
    author = "Zhou, Zhihuan and Miao, Zhuang and Bi, Sheng and Ai, Chaoqian and Zhang, Hongchao",
    title = "{What prevents resolving the Hubble tension through late-time expansion modifications?}",
    eprint = "2506.23556",
    archivePrefix = "arXiv",
    primaryClass = "astro-ph.CO",
    doi = "10.1103/bc81-3fj5",
    journal = "Phys. Rev. D",
    volume = "112",
    number = "10",
    pages = "103502",
    year = "2025"
}

@article{Pedrotti:2025ccw,
    author = "Pedrotti, Davide and Escamilla, Luis A. and Marra, Valerio and Perivolaropoulos, Leandros and Vagnozzi, Sunny",
    title = "{BAO miscalibration cannot rescue late-time solutions to the Hubble tension}",
    eprint = "2510.01974",
    archivePrefix = "arXiv",
    primaryClass = "astro-ph.CO",
    doi = "10.1103/pn9j-8whx",
    journal = "Phys. Rev. D",
    volume = "113",
    number = "4",
    pages = "043507",
    year = "2026"
}

@article{DiValentino:2020vvd,
    author = "Di Valentino, Eleonora and others",
    title = "{Cosmology Intertwined III: $f \sigma_8$ and $S_8$}",
    eprint = "2008.11285",
    archivePrefix = "arXiv",
    primaryClass = "astro-ph.CO",
    reportNumber = "FERMILAB-PUB-20-495-AE",
    doi = "10.1016/j.astropartphys.2021.102604",
    journal = "Astropart. Phys.",
    volume = "131",
    pages = "102604",
    year = "2021"
}

@article{Pantos:2026koc,
    author = "Pantos, Ioannis and Perivolaropoulos, Leandros",
    title = "{Status of the $S_8$ Tension: A 2026 Review of Probe Discrepancies}",
    eprint = "2602.12238",
    archivePrefix = "arXiv",
    primaryClass = "astro-ph.CO",
    doi = "10.1016/j.dark.2026.102286",
    journal = "Phys. Dark Univ.",
    volume = "52",
    pages = "102286",
    year = "2026"
}

@article{Hill:2020osr,
    author = "Hill, J. Colin and McDonough, Evan and Toomey, Michael W. and Alexander, Stephon",
    title = "{Early dark energy does not restore cosmological concordance}",
    eprint = "2003.07355",
    archivePrefix = "arXiv",
    primaryClass = "astro-ph.CO",
    doi = "10.1103/PhysRevD.102.043507",
    journal = "Phys. Rev. D",
    volume = "102",
    number = "4",
    pages = "043507",
    year = "2020"
}

@article{Verde:2019ivm,
    author = "Verde, L. and Treu, T. and Riess, A. G.",
    title = "{Tensions between the Early and the Late Universe}",
    eprint = "1907.10625",
    archivePrefix = "arXiv",
    primaryClass = "astro-ph.CO",
    doi = "10.1038/s41550-019-0902-0",
    journal = "Nature Astron.",
    volume = "3",
    pages = "891",
    month = "7",
    year = "2019"
}

@article{DiValentino:2020zio,
    author = "Di Valentino, Eleonora and others",
    title = "{Snowmass2021 - Letter of interest cosmology intertwined II: The hubble constant tension}",
    eprint = "2008.11284",
    archivePrefix = "arXiv",
    primaryClass = "astro-ph.CO",
    reportNumber = "FERMILAB-PUB-21-590-PPD",
    doi = "10.1016/j.astropartphys.2021.102605",
    journal = "Astropart. Phys.",
    volume = "131",
    pages = "102605",
    year = "2021"
}

@article{DiValentino:2021izs,
    author = "Di Valentino, Eleonora and Mena, Olga and Pan, Supriya and Visinelli, Luca and Yang, Weiqiang and Melchiorri, Alessandro and Mota, David F. and Riess, Adam G. and Silk, Joseph",
    title = "{In the realm of the Hubble tension\textemdash{}a review of solutions}",
    eprint = "2103.01183",
    archivePrefix = "arXiv",
    primaryClass = "astro-ph.CO",
    reportNumber = "IPPP/20/108",
    doi = "10.1088/1361-6382/ac086d",
    journal = "Class. Quant. Grav.",
    volume = "38",
    number = "15",
    pages = "153001",
    year = "2021"
}

@article{Perivolaropoulos:2021jda,
    author = "Perivolaropoulos, Leandros and Skara, Foteini",
    title = "{Challenges for \ensuremath{\Lambda}CDM: An update}",
    eprint = "2105.05208",
    archivePrefix = "arXiv",
    primaryClass = "astro-ph.CO",
    doi = "10.1016/j.newar.2022.101659",
    journal = "New Astron. Rev.",
    volume = "95",
    pages = "101659",
    year = "2022"
}

@article{Shah:2021onj,
    author = "Shah, Paul and Lemos, Pablo and Lahav, Ofer",
    title = "{A buyer\textquoteright{}s guide to the Hubble constant}",
    eprint = "2109.01161",
    archivePrefix = "arXiv",
    primaryClass = "astro-ph.CO",
    doi = "10.1007/s00159-021-00137-4",
    journal = "Astron. Astrophys. Rev.",
    volume = "29",
    number = "1",
    pages = "9",
    year = "2021"
}

@article{Abdalla:2022yfr,
    author = "Abdalla, Elcio and others",
    title = "{Cosmology intertwined: A review of the particle physics, astrophysics, and cosmology associated with the cosmological tensions and anomalies}",
    eprint = "2203.06142",
    archivePrefix = "arXiv",
    primaryClass = "astro-ph.CO",
    reportNumber = "FERMILAB-CONF-22-192-SCD",
    doi = "10.1016/j.jheap.2022.04.002",
    journal = "JHEAp",
    volume = "34",
    pages = "49--211",
    year = "2022"
}

@article{DiValentino:2022fjm,
    author = "Di Valentino, Eleonora",
    title = "{Challenges of the Standard Cosmological Model}",
    doi = "10.3390/universe8080399",
    journal = "Universe",
    volume = "8",
    number = "8",
    pages = "399",
    year = "2022"
}

@article{Hu:2023jqc,
    author = "Hu, Jian-Ping and Wang, Fa-Yin",
    title = "{Hubble Tension: The Evidence of New Physics}",
    eprint = "2302.05709",
    archivePrefix = "arXiv",
    primaryClass = "astro-ph.CO",
    doi = "10.3390/universe9020094",
    journal = "Universe",
    volume = "9",
    number = "2",
    pages = "94",
    year = "2023"
}

@article{Verde:2023lmm,
    author = {Verde, Licia and Sch\"oneberg, Nils and Gil-Mar\'\i{}n, H\'ector},
    title = "{A Tale of Many H0}",
    eprint = "2311.13305",
    archivePrefix = "arXiv",
    primaryClass = "astro-ph.CO",
    doi = "10.1146/annurev-astro-052622-033813",
    journal = "Ann. Rev. Astron. Astrophys.",
    volume = "62",
    number = "1",
    pages = "287--331",
    year = "2024"
}

@article{CosmoVerseNetwork:2025alb,
    author = "Di Valentino, Eleonora and others",
    collaboration = "CosmoVerse Network",
    title = "{The CosmoVerse White Paper: Addressing observational tensions in cosmology with systematics and fundamental physics}",
    eprint = "2504.01669",
    archivePrefix = "arXiv",
    primaryClass = "astro-ph.CO",
    doi = "10.1016/j.dark.2025.101965",
    journal = "Phys. Dark Univ.",
    volume = "49",
    pages = "101965",
    year = "2025"
}

@article{Kamionkowski:2022pkx,
    author = "Kamionkowski, Marc and Riess, Adam G.",
    title = "{The Hubble Tension and Early Dark Energy}",
    eprint = "2211.04492",
    archivePrefix = "arXiv",
    primaryClass = "astro-ph.CO",
    doi = "10.1146/annurev-nucl-111422-024107",
    journal = "Ann. Rev. Nucl. Part. Sci.",
    volume = "73",
    pages = "153--180",
    year = "2023"
}

@article{Svrcek:2006yi,
    author = "Svrcek, Peter and Witten, Edward",
    title = "{Axions In String Theory}",
    eprint = "hep-th/0605206",
    archivePrefix = "arXiv",
    reportNumber = "SLAC-PUB-11894",
    doi = "10.1088/1126-6708/2006/06/051",
    journal = "JHEP",
    volume = "06",
    pages = "051",
    year = "2006"
}

@article{Cicoli:2012sz,
    author = "Cicoli, Michele and Goodsell, Mark and Ringwald, Andreas",
    title = "{The type IIB string axiverse and its low-energy phenomenology}",
    eprint = "1206.0819",
    archivePrefix = "arXiv",
    primaryClass = "hep-th",
    reportNumber = "DESY-12-058, CERN-PH-TH-2012-153",
    doi = "10.1007/JHEP10(2012)146",
    journal = "JHEP",
    volume = "10",
    pages = "146",
    year = "2012"
}

@article{Visinelli:2018utg,
    author = "Visinelli, Luca and Vagnozzi, Sunny",
    title = "{Cosmological window onto the string axiverse and the supersymmetry breaking scale}",
    eprint = "1809.06382",
    archivePrefix = "arXiv",
    primaryClass = "hep-ph",
    reportNumber = "NORDITA-2018-051",
    doi = "10.1103/PhysRevD.99.063517",
    journal = "Phys. Rev. D",
    volume = "99",
    number = "6",
    pages = "063517",
    year = "2019"
}

@article{Cicoli:2023opf,
    author = "Cicoli, Michele and Conlon, Joseph P. and Maharana, Anshuman and Parameswaran, Susha and Quevedo, Fernando and Zavala, Ivonne",
    title = "{String cosmology: From the early universe to today}",
    eprint = "2303.04819",
    archivePrefix = "arXiv",
    primaryClass = "hep-th",
    doi = "10.1016/j.physrep.2024.01.002",
    journal = "Phys. Rept.",
    volume = "1059",
    pages = "1--155",
    year = "2024"
}

@article{Blas:2011rf,
    author = "Blas, Diego and Lesgourgues, Julien and Tram, Thomas",
    title = "{The Cosmic Linear Anisotropy Solving System (CLASS) II: Approximation schemes}",
    eprint = "1104.2933",
    archivePrefix = "arXiv",
    primaryClass = "astro-ph.CO",
    reportNumber = "CERN-PH-TH-2011-082, LAPTH-010-11",
    doi = "10.1088/1475-7516/2011/07/034",
    journal = "JCAP",
    volume = "07",
    pages = "034",
    year = "2011"
}

@article{Poulin:2018dzj,
    author = "Poulin, Vivian and Smith, Tristan L. and Grin, Daniel and Karwal, Tanvi and Kamionkowski, Marc",
    title = "{Cosmological implications of ultralight axionlike fields}",
    eprint = "1806.10608",
    archivePrefix = "arXiv",
    primaryClass = "astro-ph.CO",
    doi = "10.1103/PhysRevD.98.083525",
    journal = "Phys. Rev. D",
    volume = "98",
    number = "8",
    pages = "083525",
    year = "2018"
}

@article{Murgia:2020ryi,
    author = "Murgia, Riccardo and Abell{\'a}n, Guillermo F. and Poulin, Vivian",
    title = "{Early dark energy resolution to the Hubble tension in light of weak lensing surveys and lensing anomalies}",
    eprint = "2009.10733",
    archivePrefix = "arXiv",
    primaryClass = "astro-ph.CO",
    doi = "10.1103/PhysRevD.103.063502",
    journal = "Phys. Rev. D",
    volume = "103",
    number = "6",
    pages = "063502",
    year = "2021"
}

@article{Planck:2019nip,
    author = "Aghanim, N. and others",
    collaboration = "Planck",
    title = "{Planck 2018 results. V. CMB power spectra and likelihoods}",
    eprint = "1907.12875",
    archivePrefix = "arXiv",
    primaryClass = "astro-ph.CO",
    doi = "10.1051/0004-6361/201936386",
    journal = "Astron. Astrophys.",
    volume = "641",
    pages = "A5",
    year = "2020"
}

@article{Planck:2018lbu,
    author = "Aghanim, N. and others",
    collaboration = "Planck",
    title = "{Planck 2018 results. VIII. Gravitational lensing}",
    eprint = "1807.06210",
    archivePrefix = "arXiv",
    primaryClass = "astro-ph.CO",
    doi = "10.1051/0004-6361/201833886",
    journal = "Astron. Astrophys.",
    volume = "641",
    pages = "A8",
    year = "2020"
}

@article{AtacamaCosmologyTelescope:2025blo,
    author = "Louis, Thibaut and others",
    collaboration = "Atacama Cosmology Telescope",
    title = "{The Atacama Cosmology Telescope: DR6 power spectra, likelihoods and {\ensuremath{\Lambda}}CDM parameters}",
    eprint = "2503.14452",
    archivePrefix = "arXiv",
    primaryClass = "astro-ph.CO",
    reportNumber = "FERMILAB-PUB-25-0071-PPD",
    doi = "10.1088/1475-7516/2025/11/062",
    journal = "JCAP",
    volume = "11",
    pages = "062",
    year = "2025"
}

@article{Lynch:2024gmp,
    author = "Lynch, Gabriel P. and Knox, Lloyd and Chluba, Jens",
    title = "{Reconstructing the recombination history by combining early and late cosmological probes}",
    eprint = "2404.05715",
    archivePrefix = "arXiv",
    primaryClass = "astro-ph.CO",
    doi = "10.1103/PhysRevD.110.063518",
    journal = "Phys. Rev. D",
    volume = "110",
    number = "6",
    pages = "063518",
    year = "2024"
}

@article{An:2023buh,
    author = "An, Rui and Gluscevic, Vera",
    title = "{Reconstructing the early-Universe expansion and thermal history}",
    eprint = "2310.17195",
    archivePrefix = "arXiv",
    primaryClass = "astro-ph.CO",
    doi = "10.1103/PhysRevD.109.023534",
    journal = "Phys. Rev. D",
    volume = "109",
    number = "2",
    pages = "023534",
    year = "2024"
}

@article{Mirpoorian:2024fka,
    author = "Mirpoorian, Seyed Hamidreza and Jedamzik, Karsten and Pogosian, Levon",
    title = "{Modified recombination and the Hubble tension}",
    eprint = "2411.16678",
    archivePrefix = "arXiv",
    primaryClass = "astro-ph.CO",
    doi = "10.1103/PhysRevD.111.083519",
    journal = "Phys. Rev. D",
    volume = "111",
    number = "8",
    pages = "083519",
    year = "2025"
}

@article{Agrawal:2019lmo,
    author = "Agrawal, Prateek and Cyr-Racine, Francis-Yan and Pinner, David and Randall, Lisa",
    title = "{Rock {\textquoteleft}n{\textquoteright} roll solutions to the Hubble tension}",
    eprint = "1904.01016",
    archivePrefix = "arXiv",
    primaryClass = "astro-ph.CO",
    doi = "10.1016/j.dark.2023.101347",
    journal = "Phys. Dark Univ.",
    volume = "42",
    pages = "101347",
    year = "2023"
}

@article{Lin:2019qug,
    author = "Lin, Meng-Xiang and Benevento, Giampaolo and Hu, Wayne and Raveri, Marco",
    title = "{Acoustic Dark Energy: Potential Conversion of the Hubble Tension}",
    eprint = "1905.12618",
    archivePrefix = "arXiv",
    primaryClass = "astro-ph.CO",
    doi = "10.1103/PhysRevD.100.063542",
    journal = "Phys. Rev. D",
    volume = "100",
    number = "6",
    pages = "063542",
    year = "2019"
}
